      \documentstyle[12pt,graphicx]{article}                
\begin{document}
\baselineskip 18pt
\def\today{\ifcase\month\or
 January\or February\or March\or April\or May\or June\or
 July\or August\or September\or October\or November\or December\fi
 \space\number\day, \number\year}
\def\thebibliography#1{\section*{References\markboth
 {References}{References}}\list
 {[\arabic{enumi}]}{\settowidth\labelwidth{[#1]}
 \leftmargin\labelwidth
 \advance\leftmargin\labelsep
 \usecounter{enumi}}
 \def\newblock{\hskip .11em plus .33em minus .07em}
 \sloppy
 \sfcode`\.=1000\relax}
\let\endthebibliography=\endlist
\def\lsim{\ ^<\llap{$_\sim$}\ }
\def\gsim{\ ^>\llap{$_\sim$}\ }
\def\r2{\sqrt 2}
\def\beq{\begin{equation}}
\def\eeq{\end{equation}}
\def\beqn{\begin{eqnarray}}
\def\eeqn{\end{eqnarray}}
\def\rmuu{\gamma^{\mu}}
\def\rmud{\gamma_{\mu}}
\def\PL{{1-\gamma_5\over 2}}
\def\PR{{1+\gamma_5\over 2}}
\def\sinW2{\sin^2\theta_W}
\def\AEM{\alpha_{EM}}
\def\mul{M_{\tilde{u} L}^2}
\def\mur{M_{\tilde{u} R}^2}
\def\mdl{M_{\tilde{d} L}^2}
\def\mdr{M_{\tilde{d} R}^2}
\def\mz2{M_{z}^2}
\def\c2b{\cos 2\beta}
\def\au{A_u}
\def\ad{A_d}
\def\cob{\cot \beta}
\def\v#1{v_#1}
\def\tb{\tan\beta}
\def\epem{$e^+e^-$}
\def\KK{$K^0$-$\bar{K^0}$}
\def\wi{\omega_i}
\def\xj{\chi_j}
\def\Wmu{W_\mu}
\def\Wnu{W_\nu}
\def\m#1{{\tilde m}_#1}
\def\mH{m_H}
\def\mw#1{{\tilde m}_{\omega #1}}
\def\mx#1{{\tilde m}_{\chi^{0}_#1}}
\def\mc#1{{\tilde m}_{\chi^{+}_#1}}
\def\mwi{{\tilde m}_{\omega i}}
\def\mxi{{\tilde m}_{\chi^{0}_i}}
\def\mci{{\tilde m}_{\chi^{+}_i}}
\def\mz{M_z}
\def\sw{\sin\theta_W}
\def\cw{\cos\theta_W}
\def\cb{\cos\beta}
\def\sb{\sin\beta}
\def\rwi{r_{\omega i}}
\def\rxj{r_{\chi j}}
\def\rfp{r_f'}
\def\Kik{K_{ik}}
\def\Fq2{F_{2}(q^2)}
\def\tw{\tan\theta_W}
\def\sec2w{sec^2\theta_W}

\begin{titlepage}

\begin{center}
{\large {\bf CP Violation and the Muon Anomaly in $N=1$ Supergravity}}\\

\vskip 0.5 true cm
\vspace{2cm}
\renewcommand{\thefootnote}
{\fnsymbol{footnote}}
 Tarek Ibrahim$^a$ and Pran Nath$^b$  
\vskip 0.5 true cm
\end{center}
\noindent
{a. Department of  Physics, Faculty of Science,
University of Alexandria,}\\
{ Alexandria, Egypt}\\ 
{b. Department of Physics, Northeastern University,
Boston, MA 02115-5005, USA } \\

\vskip 1.0 true cm

\centerline{\bf Abstract}
\medskip
The one loop supersymmetric electro-weak correction 
to the anomalous magnetic moment of the muon is derived in the
minimal $N=1$ supergravity unification with two CP violating phases.
A numerical analysis of the CP violating effects on $g_{\mu}-2$ is 
carried out with the cancellation mechanism to guarantee the 
satisfaction of the experimental limits on the electric dipole moments
 of the electron and on the neutron. It is found that the
effects of the CP phases can significantly affect the supersymmetric 
electro-weak correction to $g_{\mu}-2$,
and that the numerical size of such a correction can be as large or 
larger than the Standard Model electro-weak correction to  $g_{\mu}-2$. 
These results are of import for the new Brookhaven experiment which is 
expected to increase the sensitivity of the 
$g_{\mu}-2$ measurements by a factor of 20 and allow for a test 
of the electro-weak correction in the near future.
\end{titlepage}

\section{Introduction}
 Supersymmetric theories contain a large number of CP violating
 phases which  arise from the soft supersymmetry (SUSY) breaking sector
 of the theory and contribute to the electric dipole moment (EDM) of 
 the quarks and the leptons. Currently there exist stringent limits
 on the neutron\cite{harris} and on the electron\cite{commins}
  EDM. Thus CP violations in supersymmetric theories 
  is severely constrained by experiment. To satisfy these constraints 
  it has generally been assumed that the CP violating phases are 
  small\cite{ellis, wein}. However, small phases constitute a fine tuning
  and an alternative possibility suggested is that that the CP violating
  phases can be large O(1) and the EDM constraints 
  could be satisfied by the choice of a heavy spectrum\cite{na}.  However, 
  for CP phases O(1) the satisfaction of the EDM constraints may require 
  the SUSY spectrum to lie in the several TeV region thus putting the 
  spectrum even beyond the reach of the Large Hadron Collider (LHC). 
 More recently a third possibility has been proposed, and that is 
 of internal cancellations among various contributions to the 
 electron and the neutron EDMs\cite{in1}, and there have been
 further developments of this idea\cite{in2,fo,bgk,bartl,prs}. 
 Since the cancellation mechanism allows for the
 possibility of large CP violating phases, it is of considerable 
 interest to explore the effects of such large phases on
 low energy physics and several studies exploring the effects of large
 phases have recently been reported. These include the effects of large
 CP phases on dark matter\cite{ffo,cin}, on low energy 
 phenomena\cite{kane,pw,barger,more},
 as well as other SUSY 
 phenomena\cite{pilaftsis,demir,babu,bk,everett,accomando}.

 In this paper we investigate the effects of CP violation on the 
 supersymmetric electro-weak contributions to $g_{\mu}-2$.
  This analysis extends the previous analyses of supersymmtric 
  electro-weak contributions  without the inclusion of the 
  CP violating effects\cite{grifols,kos}. This investigation is timely since
 the Brookhaven experiment E821 has started collecting data and
 in the near future will improve the sensitivity of the $g_{\mu}-2$
 measurements to allow a test of the Standard Model electro-weak
 contribution\cite{bnl1}. 
 It is already known that the supersymmetric electro-weak
 contributions to $g_{\mu}-2$ can be as large or 
 larger\cite{kos,lopez,chatto} than the
 Standard Model electro-weak contribution\cite{fuji}
and it is thus of interest to investigate the effects of large
CP violating phases on the supersymmetric muon anomaly.

We begin by exhibiting the SUSY breaking sector of the 
CP violating phases relevant for our case.  It is 
given by 
\beqn
V_{SB} & &={m_1^2|H_1|^2+m_2^2|H_2|^2 -
  [B \mu\epsilon_{ij} H_1^i H_2^j+H.c.]} 
{+m_{\tilde{L}}^2[\tilde{\nu}_{\mu}^*\tilde{\nu}_{\mu}+
\tilde{\mu}_{L}^*\tilde{\mu}_{L}]
+m_{\tilde{R}}^2 \tilde{\mu}_{R}^*\tilde{\mu}_{R}}\nonumber\\
& &
{+\frac{g m_0}{\r2 m_W} \epsilon_{ij}
[\frac{m_{\mu} A_{\mu}}{\cb} H_1^i \tilde{l}_{L}^j
\tilde{\mu}_{R}^* +H.c. ]}
+\frac{1}{2}[{\m2 \bar{\tilde{W}}^a \tilde{W}^a+\m1 \bar{\tilde{B}} \tilde{B}]
   +\Delta V_{SB}}
\eeqn
where  $\tilde{l}_{L}$ is the $SU(2)_L$ smuon doublet, 
$tan\beta=|<H_2>/<H_1>|$ where $H_1$ gives mass to the muon.
The quantities $A_{\mu}$, $\mu$, and $B$ are in general complex. 

In this analysis we shall limit ourselves to the framework of 
the minimal supergravity model\cite{chams}. In the minimal
supergravity framework (mSUGRA)
 the soft SUSY breaking is characterized by
the parameters
 $m_0$,  $m_{\frac{1}{2}}$,  $A_0$,  $\tb$,  $\theta_{\mu0}$ and $\alpha_{A0}$,
where $m_0$ is the universal scalar mass at the GUT scale, 
$m_{\frac{1}{2}}$ is the universal gauginos mass at the GUT scale, 
$A_0$ is the universal trilinear coupling at the GUT scale, 
 $\theta_{\mu0}$ is the phase of $\mu_0$ at the GUT scale, 
 and $\alpha_{A0}$ is the phase of $A_0$.   
 In the analysis we use one-loop renormalization
group equations (RGEs) for the evolution of the soft SUSY breaking parameters
and for the parameter $\mu$, and two-loop RGEs for the gauge and 
Yukawa couplings.  The phase of $\mu$ does not run because it cancels out
of the one loop renormalization group equation of $\mu$. However the
magnitude and the phase of $A_{\mu}$ do evolve.
Thus while the phase of $A_{\mu}$ is modified from $\alpha_{A_{\mu 0}}$
at the GUT scale to its value $\alpha_{A_{\mu }}$ at the electro-weak scale,
the phase of $\mu$ is unaffected at the one loop level, i.e.,
$\theta_{\mu}= \theta_{\mu0}$.

 The outline of the rest of the paper is as follows: 
 In Sec.2 we derive a
 general formula for the contribution to 
  $a_f=\frac{g_f-2}{2}$ in the presence of CP violating phases. 
  In Sec.3 we compute the 
 supersymmetric electro-weak corrections to $a_{\mu}$ from the
 chargino exchange and in Sec. 4 we compute the supersymmetric 
 electro-weak corrections to $a_{\mu}$ from the neutralino 
 exchange.  A discussion of these results is given in Sec. 5 and
 a numerical analysis of the effects of CP violating 
 phases is given in Sec. 6. We summarize our results in Sec. 7. 

\section{g-2 Calculation with CP Violation in SUSY}

In this section we derive  a general formula  for the 
contribution to $a_{\mu}$ for an interaction with CP
violating phases which would be typical of the interactions
that we will encounter in Secs. 3 and 4. 
For a theory of a fermion $\psi_f$ 
of mass $m_f$ interacting with other heavy
 fermions $\psi_i$'s and heavy scalars $\phi_k$'s with
masses $m_i$ and $m_k$, the 
interaction that contains CP violation is in general given by
\beq
-{\cal L}_{int}= \sum_{ik}\bar{\psi_f}(K_{ik}\PL +L_{ik}\PR)
\psi_i \phi_k + H.c. 
\eeq
Here $\cal L$ violates CP invariance
 iff $ Im(K_{ik}L_{ik}^*)$ is different from zero. 
The one loop contribution to $a_f$ is given by
\beq
a_f=a_f^1+a_f^2
\eeq
where $a_f^1$ and $a_f^2$ are coming from Fig. 1(a)
 and Fig. 1(b) respectively. 
$a_f^1$ is a sum of two terms:  $a_f^1=a_f^{11}+a_f^{12}$ where
\beq
a_f^{11}=\sum_{ik} \frac{m_f}{16 \pi^2 m_i}Re(K_{ik}L_{ik}^*)
F_1(\frac{m_k^2}{m_i^2})
\eeq
and
\beq
F_1(x)=\frac{1}{(x-1)^3}(1-x^2+2x lnx)
\eeq
and where 
\beq
 a_f^{12}=\sum_{ik} \frac{m_f^2}{96 \pi^2 m_i^2}(|K_{ik}|^2+|L_{ik}|^2)
F_2(\frac{m_k^2}{m_i^2})
\eeq
and 
\beq
F_2(x)=\frac{1}{(x-1)^4}(-x^3+6x^2-3x-2-6x lnx).
\eeq
Similarly, $a_f^2$ also consists of two terms:
 $a_f^2=a_f^{21}+a_f^{22}$ where 
\beq
a_f^{21}=-\sum_{ik}\frac{m_f}{16 \pi^2 m_i}Re(K_{ik}L_{ik}^*)
F_3(\frac{m_k^2}{m_i^2})
\eeq
and 
\beq
F_3(x)=\frac{1}{(x-1)^3}(3x^2-4x+1-2x^2 lnx)
\eeq
and where 
\beq
a_f^{22}=\sum_{ik}\frac{m_f^2}{96\pi^2 m_i^2}
(|K_{ik}|^2+|L_{ik}|^2)F_4(\frac{m_k^2}{m_i^2})
\eeq
and 
\beq
F_4(x)=\frac{1}{(x-1)^4}(2x^3+3x^2-6x+1-6x^2 lnx).
\eeq

\section{Chargino Contributions with CP Violating Phases}
The chargino matrix with CP violating phases is given by
\beq
M_C=\left(\matrix{\m2 & \r2 m_W  \sb \cr
	\r2 m_W \cb & |\mu| e^{i\theta_{\mu}}}
            \right)
\eeq
This matrix can be diagonalized by a biunitary transformation
$U^* M_C V^{-1}\\
=diag(\mc1,\mc2)$.  
 By looking at the muon-sneutrino-chargino
interaction:
\beqn
-{\cal L}_{\mu-\tilde{\nu}-\tilde{\chi}^{+}} & & =
{g \bar{\mu}[V_{11}P_{R}-\kappa_{\mu}U_{12}^{*}P_{L}]
\tilde{\chi}_1^{+} \tilde{\nu}} \nonumber\\
& &
{+g \bar{\mu}[V_{21}P_{R}-\kappa_{\mu}U_{22}^{*}P_{L}]
\tilde{\chi}_2^{+} \tilde{\nu}} + H.c., 
\eeqn
where $\kappa_{\mu}=\frac{m_{\mu}}{\sqrt 2 M_W \cos\beta}$,
  we find that the chargino exchange to 
$a_{\mu}$ is given by
\beq
a^{\chi^+}_{\mu}=a^{21}_{\mu}+a^{22}_{\mu}
\eeq
where
\beq
a^{21}_{\mu}=\frac{m_{\mu}\alpha_{EM}}{4\pi\sin^2\theta_W}
\frac{m_{\mu}}{\r2 m_W \cb}
\sum_{i=1}^{2}\frac{1}{M_{\chi_i^+}}Re(U^*_{i2}V^*_{i1})
F_3(\frac{M^2_{\tilde{\nu}}}{M^2_{\chi_i^+}})
\eeq
and
\beq
a^{22}_{\mu}=\frac{m^2_{\mu}\alpha_{EM}}{24\pi\sin^2\theta_W}
\sum_{i=1}^{2}\frac{1}{M^2_{\chi_i^+}}
(\frac{m^2_{\mu}}{2 m^2_W \cos^2\beta} |U_{i2}|^2+|V_{i1}|^2)
F_4(\frac{M^2_{\tilde{\nu}}}{M^2_{\chi_i^+}}).
\eeq
The phase which enters here is $\theta_{\mu}$ through the matrix 
elements of $U$ and $V$.

\section{Neutralino Contributions with CP Violating Phases}
The neutralino mass matrix $M_{\chi^0}$ is a complex symmetric matrix 
and is given by
\beq
\left(\matrix{\m1
 & 0 & -\mz\sw\cb  & \mz\sw\sb  \cr
  0  & \m2  & \mz\cw\cb & -\mz\cw\sb \cr
-\mz\sw\cb  & \mz\cw\cb  & 0 &
 -|\mu| e^{i\theta_{\mu}}\cr
\mz\sw\sb  & -\mz\cw\sb 
& -|\mu| e^{i\theta_{\mu}} & 0}
                        \right)
\eeq
 The matrix $M_{\chi^0}$ can
be diagonalized by the unitary transformation 

\beq
X^T M_{\chi^0} X={\rm diag}(\mx1, \mx2, \mx3, \mx4).
\eeq
The smuon $(mass)^2$ matrix is given by   
	 \beq
M_{\tilde{\mu}}^2=\left(\matrix{M_{\tilde{L}}^2+m{_{\mu}}^2
-M_{z}^2(\frac{1}{2}-
\sin^2\theta_W)\cos2\beta & m_{\mu}(A_{\mu}^{*}m_0-\mu \tan\beta) \cr
   	          	m_{\mu}(A_{\mu} m_0-\mu^{*} \tan\beta) & M_{\tilde R}
^2+m{_\mu}^2-M_{Z}^2  \sin^2\theta_W \cos2\beta}
		\right)
\eeq 
 This matrix is  hermitian and can be diagonalized by the unitary transformation
\beq
D_{\mu}^\dagger M_{\tilde{\mu}}^2 D_{\mu}={\rm diag}(M_{\tilde{\mu}1}^2,
              M_{\tilde{\mu}2}^2)
\eeq
The muon-smuon-neutralino interaction in the mass
diagonal basis is defined by 
\beqn
-{\cal L}_{\mu-\tilde{\mu}-\tilde{\chi}^{0}} & & = {\sum_{j=1}^{4}
\r2 \bar{\mu}[(\alpha_{\mu j}
D_{\mu 11}-\gamma_{\mu j} D_{\mu 21})P_{L}} \nonumber\\
& &
{+(\beta_{\mu j} D_{\mu 11}-\delta_{\mu j}  D_{\mu 21})P_{R}]
\tilde{\chi}_j^{0} \tilde{\mu}_1}
{+\r2 \bar{\mu}[(\alpha_{\mu j}
D_{\mu 12}-\gamma_{\mu j} D_{\mu 22})P_{L}} \nonumber\\
& &
{+(\beta_{\mu j} D_{\mu 12}-\delta_{\mu j}  D_{\mu 22})P_{R}]
\tilde{\chi}_j^{0} \tilde{\mu}_2 +H.c.}
\eeqn
where $\alpha$, $\beta$, $\gamma$ and $\delta$ are given by
\beq
\alpha_{\mu j}=\frac{gm_{\mu}X_{3,j}}{2m_W  \cos{\beta}}
\eeq
\beq
\beta_{\mu j}=eQ_{\mu}X_{1j}^{'*} +\frac{g}{\cos{\theta_W}}
 X_{2j}^{'*}(T_{3\mu}-Q_{\mu}\sin^{2}{\theta_W})
\eeq
\beq
\gamma_{\mu j}=eQ_{\mu}X_{1j}^{'} -\frac{gQ_{\mu}\sin^{2}{\theta_W}}
{\cos{\theta_W}}  X_{2j}^{'}
\eeq
\beq
\delta_{\mu j}=-\frac{gm_{\mu}X_{3,j}^*}{2m_W  \cos{\beta}}
\eeq
and where
\beq
X_{1j}^{'}=X_{1j} \cos{\theta_W}+X_{2j} \sin{\theta_W}
\eeq
\beq
X_{2j}^{'}=-X_{1j} \sin{\theta_W}+X_{2j} \cos{\theta_W},
\eeq
The neutralino exchange contribution to $a_{\mu}$ is given by
\beq
a^{\chi^0}_{\mu}=a^{11}_{\mu}+a^{12}_{\mu}
\eeq
where
\beq
a^{11}_{\mu}=\frac{m_{\mu}\alpha_{EM}}{4\pi\sin^2\theta_W}
\sum_{j=1}^{4}\sum_{k=1}^{2}\frac{1}{M_{\chi_j^0}} \eta^k_{\mu j}
F_1(\frac{M^2_{\tilde{\mu_k}}}{M^2_{\chi_j^0}})
\eeq
and
\beq
a^{12}_{\mu}=\frac{m^2_{\mu}\alpha_{EM}}{24\pi\sin^2\theta_W}
\sum_{j=1}^{4}\sum_{k=1}^{2}\frac{1}{M^2_{\chi_j^0}}X^k_{\mu j}
F_2(\frac{M^2_{\tilde{\mu_k}}}{M^2_{\chi_j^0}})
\eeq
Here
\beqn
\eta^k_{\mu j} & &=-\tan^2\theta_W Re(X^2_{1j}D_{1k}^*D_{2k})
-\tan\theta_W Re(X_{2j}X_{1j}D_{1k}^*D_{2k})\nonumber\\
&&
+\frac{m_{\mu}\tan\theta_W}{M_W \cos\beta}|D_{2k}|^2 
Re(X_{3j}X_{1j})
-\frac{m_{\mu}\tan\theta_W}{2 M_W \cos\beta}|D_{1k}|^2
Re(X_{3j}X_{1j})\nonumber\\ 
&&
-\frac{m_{\mu}}{2 M_W \cos\beta}|D_{1k}|^2
Re(X_{3j}X_{2j})
+\frac{m^2_{\mu}}{2 M^2_W \cos^2\beta}Re(X^2_{3j}D_{2k}^*D_{1k})
\eeqn
and
\beqn
X^k_{\mu j}&&=\frac{m^2_{\mu}}{2 M^2_W \cos^2\beta}|X_{3j}|^2 \nonumber\\
&&
+\frac{1}{2}\tan^2\theta_W |X_{1j}|^2
(|D_{1k}|^2+4|D_{2k}|^2)
+\frac{1}{2} |X_{2j}|^2|D_{1k}|^2\nonumber\\
&&
+\tan\theta_W |D_{1k}|^2 Re(X_{1j}X_{2j}^*)\nonumber\\
&&
+\frac{m_{\mu}\tan\theta_W}{M_W \cos\beta}
Re(X_{3j}X_{1j}^*D_{1k}D_{2k}^*)
-\frac{m_{\mu}}{M_W\cos\beta}Re(X_{3j}X_{2j}^*D_{1k}D_{2k}^*)
\eeqn
 The matrix elements of $X$ carry  the phase of $\mu$ 
and the matrix elements of $D$ carry  both the phase of 
$\mu$ and the phase  of the trilinear parameter $A_{\mu}$
where $A_{\mu}$ is the renormalization group evolved value of
$A_{\mu_0}$ at the  Z-scale. 

\section{Discussion of Results}
It is interesting to consider the supersymmetric limit  
 of our results when the soft susy breaking terms vanish. In this
  limit Eq.(14) which arises from the chargino exchange gives a
  contribution which is equal in magnitude and opposite in sign to the
   contribution from the W exchange. Thus we find that in the
   supersymmetric limit 
   \beq
   a_{\mu}^W+a_{\mu}^{\chi^+}=0
   \eeq
   Similarly taking the  supersymmetric limit of Eq.(28) we
   find that the massive modes
 neutralino exchange contribution is equal in magnitude
   and opposite in sign to the Z boson exchange contribution so that
   
   \beq
     a_{\mu}^Z+a_{\mu}^{\chi^0}(massive~modes)=0
   \eeq
   This is what one expects on general grounds\cite{fr,bg} and
   our explicit evaluations satisfy Eqs.(33) and (34).  The proof
   of Eqs.(33) and (34) is given in Appendix A.  
   A result similar to Eq.(34) holds for the massless modes 
   but its proof requires extension of the results of Sec.2 to
   include $m_f$ corrections in the loop integrals.
   This extension will be discussed elsewhere.   
  	
	Next we discuss the limit of vanishing CP violating phases.
	In this limit the unitary 
	matrices U and V becomes orthogonal matrices. Using the
	notation 
\beq
V^{-1}\rightarrow O_1, U^*\rightarrow O_2^T
\eeq	
where $O_1$ and $O_2$ are orthogonal matrices, 
 the chargino contributions take on the form 
\beq
a^{21}_{\mu}=\frac{m_{\mu}\alpha_{EM}}{4\pi\sin^2\theta_W}
\frac{m_{\mu}}{\r2 m_W \cb}
\sum_{i=1}^{2}\frac{1}{M_{\chi_i^+}}O_{22i}O_{1i1}^T
F_3(\xi_{\nu i})
\eeq
and
\beq
a^{22}_{\mu}=\frac{m^2_{\mu}\alpha_{EM}}{24\pi\sin^2\theta_W}
\sum_{i=1}^{2}\frac{1}{M^2_{\chi_i^+}}
(\frac{m^2_{\mu}}{2 m^2_W \cos^2\beta} (O_{22i})^2+(O_{1i1}^T)^2)
F_4(\xi_{\nu i})
\eeq
where $\xi_{\nu i}= {M^2_{\tilde{\nu}}}/{M^2_{\chi_i^+}}$.
The neutralino exchange contributions in the CP violating limit 
can similarly be gotten from Eqs.(28-32) by the replacement
\beq
X\rightarrow O, ~~D\rightarrow S
\eeq
where O and S are orthogonal matrices. 
Our results for the chargino and neutralino contributions go to
the  result of the previous works \cite{kos} in the vanishing CP phase
 limit considered above.

\section{Analysis of CP Violating Effects}

Before discussing the effects of CP violating phases on the 
supersymmetric contributions to $a_{\mu}$ we summarize briefly
the current experimental and theoretical situation regarding $a_{\mu}$.
The most accurate determination of $a_{\mu}$ is from the CERN
experiment\cite{bailey} which gives a value of $a_{\mu}^{exp}=
11659230(84)\times 10^{-10}$ while the Standard Model determination
including $\alpha^5$ QED contributions\cite{kino}, hadronic vacuum 
polarization\cite{davier} and light by light 
hadronic contributions\cite{hayakawa}, and the complete
two loop standard Model electro-weak contributions\cite{czar} is 
$a_{\mu}^{SM}=11659162(6.5)\times 10^{-10}$.  Here  
essentially the entire error shown in parenthesis comes from the
  hadronic sector. It is expected that the new Brookhaven $g_{\mu}$ 
  experiment\cite{bnl1,bnl2}
will improve by a factor of 20 the determination of $a_{\mu}$ 
over the previous $a_{\mu}$ measurement\cite{bailey}, i.e., the
error in the experimental determination of $a_{\mu}$ is expected
to go down to $4\times 10^{-10}$. This means that even
with no further reduction in the hadronic error the new $g_{\mu}$
experiment will be able to test the Standard Model electro-weak 
corrections which contribute an amount\cite{czar} $a_{\mu}^{EW}(SM)$=
$15.1(0.4)\times 10^{-10}$. 
However, it has
 been pointed out that the supersymmetric electro-weak corrections
 can be as large or larger than the Standard Model electro-weak 
 corrections, and thus the new $g_{\mu}$ experiment will also  probe 
supersymmetry\cite{kos,lopez,chatto}. 
In this context it is important to know  how
large the CP violating effects are on the supersymmetric electro-weak
anomaly. 

 Previous analyses of $g_{\mu}$ in supersymmetry did not consider
  the effects of CP violating phases because the effects
 of such phases were expected  to be generally small due to the
 electric dipole moment (EDM) constraints. As mentioned in Sec.1
  in the conventional 
 scenarios the current experimental constraints on the electron
 EDM ($d_e$) and on the neutron EDM ($d_n$) are satisfied either
 by the choice of small CP violating phases\cite{ellis,wein} 
 or by the choice of a 
 heavy mass spectrum of supersymmetric particles\cite{na}. 
 For the first case, 
 the CP violating effects are small because of the smallness
 of the  CP violating phases, while for the second the supersymmetric
 contribution to $g_{\mu}-2$  will itself be small compared to the
 Standard Model result to be of relevance. However, as also pointed out
 in Sec. 1 with the cancellation mechanism\cite{in1} one can 
 satisfy the EDM constraints with large CP violating phases 
 and not too massive a SUSY spectrum 
 and thus it
 is of relevance to examine the effects of CP violating phases on 
 $a_{\mu}$.

  For the case  of the electron EDM 
 the cancellations occur between the chargino
 and the neutralino exchange contributions while for the case of 
 the neutron EDM the cancellations can occur in a two step process.
 Thus for the neutron case, the EDM receives contributions from
 the electric dipole, the chromo-electric dipole and the purely
 gluonic dimension six operators. For the electric and the chromo-electric
  dipole operators 
  cancellations 
 can occur between the chargino, the gluino and the neutralino exchange 
 contributions. There is, however, the possibility
 of a further cancellation, and that is among
 the electric dipole, the chromo-electric dipole and the purely
 gluonic contributions. Recently, it has been pointed out
 that in addition to the above contributions certain two loop graphs may 
 also contribute significantly in some
 regions of the parameter space\cite{darwin}. In our analysis we have 
 included the effects of these contributions as well.  However, we find
 that the effect of these terms is relatively small compared to the
 other contributions.
  The regions of interest in the parameter
space are those where the cancellations among different components 
happen simultaneously for the  case of the electron EDM  and 
  of the neutron EDM  so as to satisfy
the experimental lower limits, which for the neutron is\cite{harris} 

\beq
 |d_n|< 6.3\times 10^{-26}
 \eeq
 and for the electron is\cite{commins}
 \beq
 |d_e|< 4.3\times 10^{-27}
\eeq

We discuss now the size of the CP violating  effects on $a_{\mu}^{SUSY}$.
 In Fig.2 we
exhibit the effect of the variation of the CP violating phase
$\theta_{\mu_0}$ on $a_{\mu}^{SUSY}$, without the imposition of the
EDM constraint, as a function of $\theta_{\mu_0}$. The values of the
other parameters ($m_0$, $m_{\tilde g}$, $tan\beta$, 
$\alpha_{A_0}$) for the curves (1)-(5) can be read off from Table 1 
for the cases  (1) - (5). We find that the effect of the CP violating
phase is very substantial.   A similar analysis of the effects of
the variation of the CP violating phase  $\alpha_{A_0}$ on 
$a_{\mu}^{SUSY}$, also without the imposition of the EDM constraint,
is given in Fig.3.  Again the value of the parameters other than
$\alpha_{A_0}$ for the curves labeled (1)-(5) can be read off from
Table 1. Here again the effects of the CP violating
phase $\alpha_{A_0}$  are found to be quite substantial although 
not as large as those from  $\theta_{\mu_0}$. The reason for this
discrepancy is
easily understood. In the region of the parameter space
considered the  chargino contribution is large and this contribution
is independent of $\alpha_{A_0}$ since  $\alpha_{A_0}$ does not
enter in the chargino mass matrix. Thus the   $\alpha_{A_0}$ 
dependence enters only via the smuon mass matrix,
while $\theta_{\mu_0}$ enters  via all mass matrices.

 Inclusion of 
the EDM constraint puts stringent constraints on the parameter
space of mSUGRA.  As an illustration we give in Fig.4 the plot of the
EDM of the electron and for the neutron as a function of 
 $\alpha_{A0}$ for curves (1) and (3) of Fig.3. The figure
 illustrates the simultaneous cancellation occurring for the
 electron and the neutron EDM in narrow regions of $\alpha_{A0}$
 and in these regions 
  the experimental EDM constraints can be satisfied. We note 
  the appearance of two cancellation minima in the cases considered.
  These double minima reveal the strong dependence on the phases 
  of the various terms that contribute to the EDMs. 
 The effects of CP violating phases on $a_{\mu}$
 can be significant in these domains. In Table 1 we give a 
 set of illustrative
 points where  a simultaneous cancellation in the electron and
 the neutron EDM occurs. The size of the effects of CP violating phases
 on $a_{\mu}$ can be seen from the 
 values of $a_{\mu}$ at these points and the corresponding
 four CP conserving cases in Table 2.
 A comparison of the results of Tables 1 and 2 with those of Figs.2
 and 3 shows that with the inclusion of the EDM constraints the 
 CP violating effects are much reduced for the points chosen here. 
 However, even with the 
 inclusion of the EDM constraints the CP effects on 
 $a_{\mu}$ can still be quite substantial as a comparison of
 Tables 1 and 2 exhibits. Inclusion of more than two phases 
 makes the satisfaction of the EDM constraints much easier 
 and detailed analyses show that there are 
 significant regions of the parameter space where the CP violating
 phases are large and cancellations
 occur to render the electron and the neutron EDM in conformity 
 with experiment\cite{in2,bgk}. Such regions are of considerable 
 interest in the investigations of SUSY phenomena at low energy.
 The effects of CP violating phases in these regions could  be 
 substantial. However, a quantitative discussion of these effects requires 
 inclusion of nonuniversal effects which are outside the framework
 of mSUGRA model discussed here.

\begin{center} \begin{tabular}{|c|c|c|c|c|c|}
\multicolumn{6}{c}{Table~1:  } \\
\hline
  & $\theta_{\mu_0}$& $\alpha_{A_0}$ & $d_n(10^{-26} e cm)$& 
  $d_e(10^{-27} e cm)$  & $ a_{\mu}
(\theta_{\mu_0}, \alpha_{A})(10^{-9})$\\
\hline
(1) &3.108 & $-0.2$ &5.4 & $-2.7$ & $-3.6$ \\
\hline
(2) & $3.08$ & $-0.45$& 4.86 & $-4.26$ & $-2.5$ 
\\                                                                             
\hline
(3) & 3.02 & $-1.0$&  $-3.6$ &$-3.1$ &$-1.7$  
\\                                                                            
\hline 
(4) & 3.1 & $-0.2$ & $-4.9$ &$-0.93$ & $-1.1$   \\
\hline
(5) & $3.02$ & $-1.0$& $-5.0$ & 1.1 & $-.78$  \\
\hline
\end{tabular}\\

\noindent
Table caption: The other parameters corresponding to the cases (1)-(5)
are:\\ (1) $m_0$=60, $m_{1/2}$=123, $tan\beta$= 3.5, $|A_0|$= 5.45,
(2) $m_0$=65, $m_{1/2}$=119,$tan\beta$= 2.6, $|A_0|$= 2.93,
(3) $m_0$=80, $m_{1/2}$=147,$tan\beta$= 2.6, $|A_0|$= 2.93,
(4) $m_0$=120, $m_{1/2}$=228,$tan\beta$= 3.5, $|A_0|$= 5.47,
(5) $m_0$=120, $m_{1/2}$=220, $tan\beta$=2.6, $|A_0|$= 2.93, 
where all masses are in GeV units. 

\end{center}

\begin{center} \begin{tabular}{|c|c|c|c|c|}
\multicolumn{5}{c}{Table~2:  } \\
\hline
  & ${a_{\mu}(0, 0)(10^{-9})}$& ${a_{\mu}(0, \pi)(10^{-9})}$ &
 ${a_{\mu}(\pi, 0)(10^{-9})}$&
  ${a_{\mu}(\pi, \pi)(10^{-9})}$ \\
\hline
(1) &3.25 & $4.18$ &$-3.5$ & $-2.6$  \\
\hline
(2) & $2.49$ & $3.1$& $-2.6$ & $-1.98$ 
\\
\hline
(3) & 1.5 & $1.86$&  $-1.9$ &$-1.34$ 
\\
\hline
(4) & .75 & $1.12$ & $-1.13$ &$-.75$   \\
\hline
(5) & $.62$ & $.82$& $-.89$ & $-.6$  \\
\hline
\end{tabular}\\
\noindent
Table caption: The values of $a_{\mu}$ for four CP conserving
cases with  all other parameters  the same  as in the corresponding
cases in Table 1. 

\end{center}

\section{Conclusions}
In this paper we have  derived the general one loop formula for the 
effects of CP violating phases on the anomalous magnetic moment of 
a fermion. We then specialized our analysis to the case of 
the calculation of CP violating effects on the supersymmetric muon anomaly.
Here the contributions arise from the one loop chargino 
and neutralino exchange diagrams.  The numerical analysis 
of the CP violating effects is strongly constrained by the 
experimental EDM constraints on the electron\cite{commins} and 
 on the neutron\cite{harris}.
Our analysis including these constraints shows that 
 the size of the CP violating effects is strongly dependent 
on the region of the parameter space one is in and that the CP violating
 phases can produce substantial affects on the  
supersymmetric electro-weak  contribution. We also find that the 
supersymmetric contribution to the muon anomaly in the presence of 
large CP violating phases consistent with the EDM constraints can be
as large  or larger than the Standard Model electro-weak contribution. 
These results are of interest in view of the new BNL muon g-2 
experiment which will improve the accuracy of the muon g-2 
measurement by a factor of 20 and test the electro-weak correction to
$g_{\mu}-2$.

\section{Acknowledgements} 

 This research was supported in part by NSF grant 
PHY-9901057. \\

\noindent
{\bf Appendix A: The Supersymmetric Limit}\\
 
In this appendix we exhibit the supersymmetric limit of the
chargino exchange contribution. The supersymmetric limit 
corresponds to $M_{\tilde \nu}=0$, $\tilde m_i$=0 (1,2), 
$tan\beta=1$ and $\mu$
 =0. In this limit $F_3(0)=-1$, $F_4(0)=1$, and
 the unitary matrices U and V take the values
 \beq
 U=\frac{1}{\sqrt 2}\left(\matrix{1 & 1 \cr
	-1& 1}
            \right),
 V=\frac{1}{\sqrt 2}\left(\matrix{1 & 1 \cr
	1& -1}
            \right)
\eeq
where $U^*M_CV^{-1}=diag(M_W,M_W)$. 
In this limit  $a^{21}_{\mu}$, $  a^{22}_{\mu}$ and 
the total chargino contribution  $a^{\chi^+}_{\mu}$ are given by

\beq
a^{21}_{\mu}=-\frac{ \alpha_{EM}}{4\pi\sin^2\theta_W}
\frac{m_{\mu}^2}{M_W^2}, 
a^{22}_{\mu}=\frac{ \alpha_{EM}}{24\pi\sin^2\theta_W}
\frac{m_{\mu}^2}{M_W^2}
\eeq

\beq
a^{\chi^+}_{\mu}  =-\frac{5 \alpha_{EM}}{24\pi\sin^2\theta_W}
\frac{m_{\mu}^2}{M_W^2}
\eeq
The result of Eq.(43) is to be compared with the contribution arising from
the exchange of the W boson\cite{fuji} 

\beq
a_{\mu}^W=\frac{5m_{\mu}^2 G_F}{12\pi^2\sqrt 2}
\eeq
Using $G_F=\pi \alpha_{em}/(M_W^2\sqrt 2 sin^2\theta_W)$ we then find that
the sum of the chargino and the W exchange
contributions vanish in the supersymmetric limit .

  Next we consider the neutralino exchange contribution to 
  $a_{\mu}^{\chi^0}$ 
  in the supersymmetric limit. In this limit two of the eigen-values
  of the neutralino mass matrix are zero and the other two are 
   $\pm M_Z$. However, we choose the unitary transformation 
  $X$ so that the non-vanishing  eigen-values are all positive definite, 
 i.e., 
  
  \beq
  X^T M_{\chi^0} X= diag(0,0, M_Z, M_Z)
  \eeq
  In this case the unitary matrix X takes on the form
\beq
\left(\matrix{\alpha & \beta & \frac{\sin\theta_W}{\sqrt 2}
  &  i\frac{\sin\theta_W}{\sqrt 2}   \cr
  \alpha\tw & \beta\tw  & -\frac{\cos\theta_W}{\sqrt 2}
     & -i\frac{\cos\theta_W}{\sqrt 2}   \cr
 \alpha &  -\frac{1}{2}\beta \sec2w & -\frac{1}{2} & \frac{i}{2} \cr
  \alpha & -\frac{1}{2}\beta \sec2w & \frac{1}{2} & -\frac{i}{ 2} }
                        \right).
\eeq
where 
\beq
\alpha = \frac{1}{\sqrt{3+\tan^2\theta_W}}, 
\beta=\frac{1}{\sqrt{1+\tan^2\theta_W+\frac{1}{2}sec^4\theta_W}}
\eeq
The appearance of i(=$\sqrt{-1}$) in the last column in X is to guarantee that the 
eigenvalues are all positive definite.  In the supersymmetric limit 
$\eta_{\mu j}^k$ take on the following form 
\beq
\eta_{\mu j}^1=-\frac{m_{\mu}}{\sqrt 2 M_W} Re(X_{3j}X_{2j})
-\frac{m_{\mu}}{\sqrt 2 M_W} \tan\theta_WRe(X_{3j}X_{1j})
\eeq
and 
\beq
\eta_{\mu j}^2=\frac{\sqrt 2 m_{\mu}}{M_W} \tan\theta_WRe(X_{3j}X_{1j})
\eeq
while $X_{\mu j}^k$ take on the form
\beq
X_{\mu j}^1=\frac{m^2_{\mu}}{M^2_W} |X_{3j}|^2
+\frac{1}{2}\tan^2\theta_W |X_{1j}|^2
+\frac{1}{2} |X_{2j}|^2 + \tan\theta_W Re(X_{1j}X_{2j}^*)
\eeq
\beq
X_{\mu j}^2=\frac{m^2_{\mu}}{M^2_W} |X_{3j}|^2+2\tan^2\theta_W|X_{1j}|^2
\eeq
Using the above and the limit $F_1(0)=-1$, $F_2(0)=-2$  and by
ignoring the terms of higher order of $m_{\mu}$,  one finds that
$a_{\mu}^{11}$ and $a_{\mu}^{12}$ simplify as follows:

\beq
a_{\mu}^{11}= \sum_{j=3}^4 \frac{m_{\mu}^2\alpha_{EM}}
{4\r2\pi \sin^2\theta_W M_WM_Z}
 (Re(X_{3j}X_{2j})+\tan\theta_W Re(X_{3j}X_{1j})
-2 \tan\theta_W Re(X_{3j}X_{1j}))
\eeq

\beq
a_{\mu}^{12}= -2\sum_{j=3}^4 \frac{m_{\mu}^2\alpha_{EM}}
{48\pi \sin^2\theta_W M_Z^2} (5\tan\theta_W|X_{1j}|^2+|X_{2j}|^2
+2 \tan\theta_W Re(X_{1j}X_{2j}^*))
\eeq
 Substitution of the explicit form of X from Eq.(46)
  in Eqs.(52) and (53) gives

\beq
a_{\mu}^{11}=\frac{m_{\mu}^2 G_F}{2\sqrt 2 \pi^2}(\frac{1}{2})
\eeq
\beq
a_{\mu}^{12}=-\frac{m_{\mu}^2 G_F}{2\sqrt 2 \pi^2}
(\frac{4}{3}\sin^4\theta_W-\frac{2}{3}\sin^2\theta_W+\frac{1}{6})
\eeq
and 
\beq
a_{\mu}^{\chi^0}=-\frac{m_{\mu}^2 G_F}{2\sqrt 2 \pi^2}
(\frac{4}{3}\sin^4\theta_W-\frac{2}{3}\sin^2\theta_W-\frac{1}{3})
\eeq
The result of Eq.(56) is to be compared to the Standard Model 
 Z exchange contribution\cite{fuji}  
\beq
a_{\mu}^Z=\frac{m_{\mu}^2 G_F}{2\sqrt 2 \pi^2}
(-\frac{5}{12}+\frac{4}{3}(\sin^2\theta_W-\frac{1}{4})^2)
\eeq
Thus one finds that in the supersymmetric limit the sum of the
neutralino and the Z boson exchange contributions vanish.

\newpage

\begin{figure}
\begin{center}
\includegraphics[angle=0,width=3.5in]{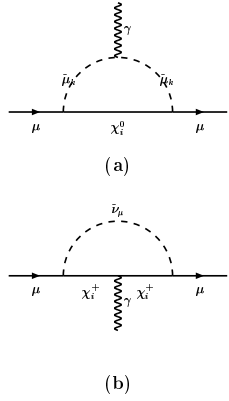}
\caption{The one loop contribution to $g_{\mu}-2$  from 
(a) neutralino exchange, and (b) chargino exchange diagrams.}
\label{fig1}
\end{center}
\end{figure}

\begin{figure}
\begin{center}
\includegraphics[angle=90,width=3.0in]{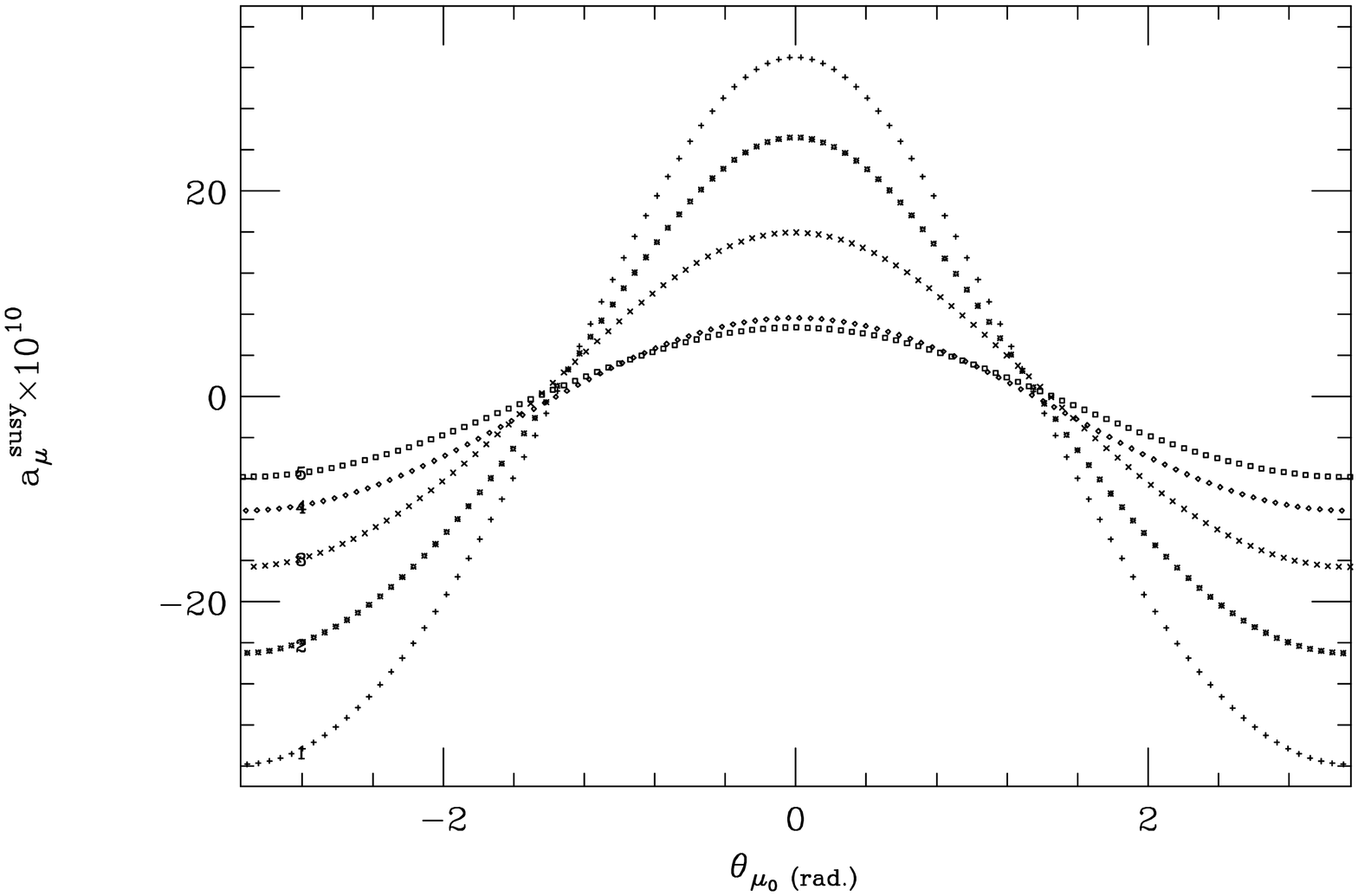}
\caption{ Plot of $a_{\mu}^{SUSY}$ as a function of the CP violating
phase $\theta_{\mu_0}$. The values of the other parameters for the
curves (1)-(5) correspond to the  cases (1)-(5) in Table 1.  }
\label{fig2}
\end{center}
\end{figure}

\begin{figure}
\begin{center}
\includegraphics[angle=90,width=3.0in]{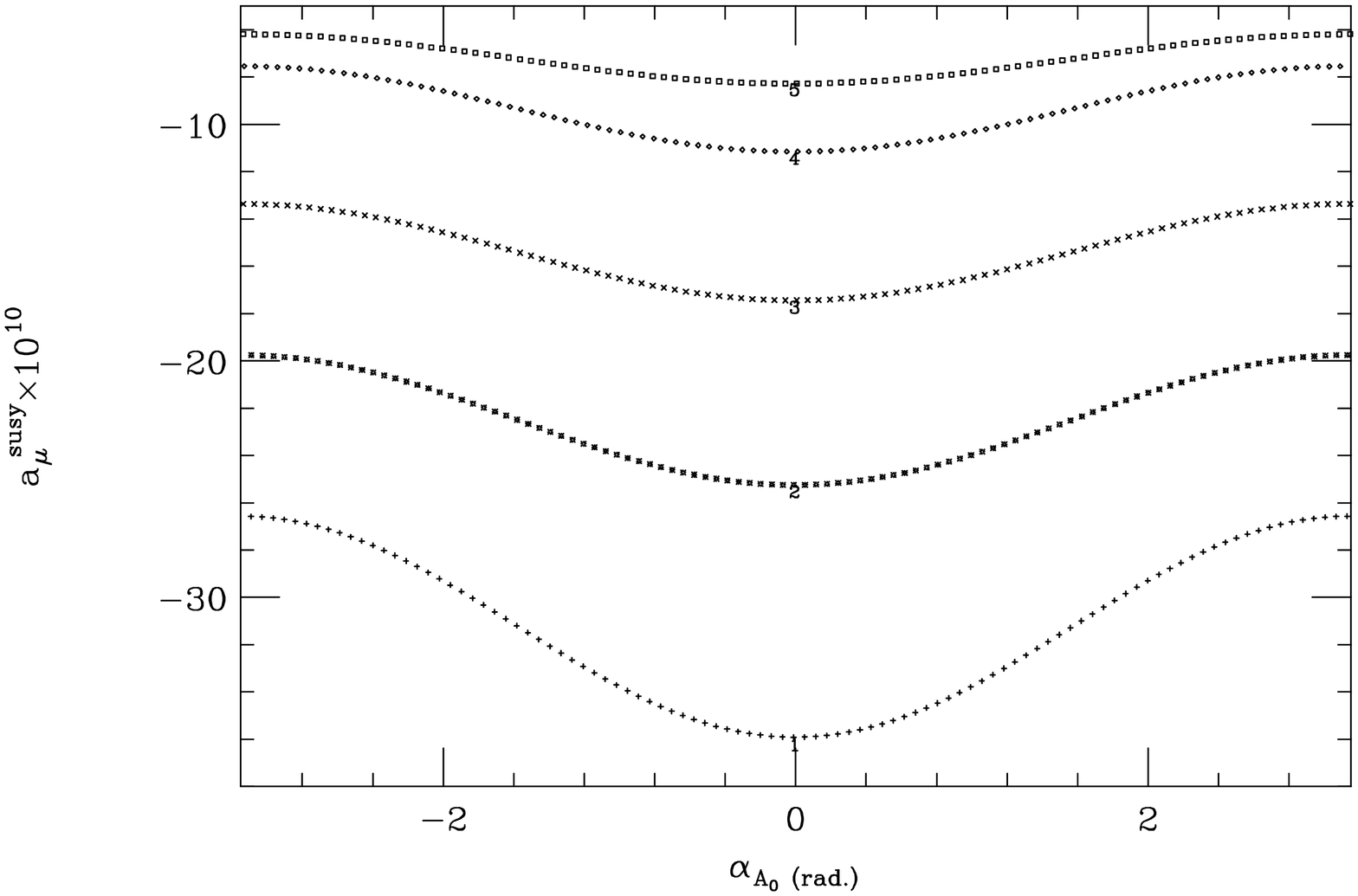}
\caption{Plot of $a_{\mu}^{SUSY}$ as a function of the CP violating
phase $\alpha_{A_0}$. The values of the other parameters for the
curves (1)-(5) correspond to the cases (1)-(5) in Table 1. }
\label{fig3}
\end{center}
\end{figure}

\begin{figure}
\begin{center}
\includegraphics[angle=90,width=3.5in]{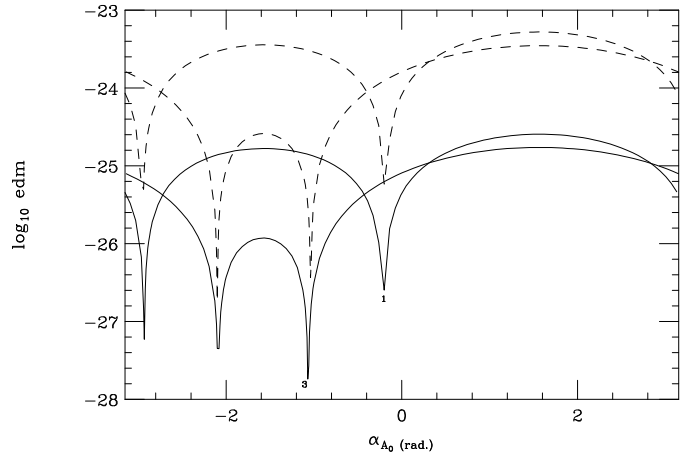}
\caption{Exhibition of the dependence of the $|EDM|$ of the electron 
(solid) and the neutron (dashed) and the cancellation 
as a function of $\alpha_{A_0}$. The curves with minima to the extreme 
left and the extreme right have other parameters corresponding to 
case (1) of Table 1, while the curves with two minima in the middle 
 have other parameters corresponding to 
case (3) of Table 1.}
\label{fig4}
\end{center}
\end{figure}

\end{document}